# Photonic non-volatile memories using phase change materials

Wolfram H.P. Pernice[*,1] and Harish Bhaskaran[2]

[1]*Institute of Nanotechnology, Karlsruhe Institute of Technology, 76133 Karlsruhe, Germany*

[2]*School of Engineering, University of Exeter, Exeter EX4 4QF, UK*

**We propose an all-photonic, non-volatile memory and processing element based on phase-change thin-films deposited onto nanophotonic waveguides. Using photonic microring resonators partially covered with $Ge_2Sb_2Te_5$ (GST) multi-level memory operation in integrated photonic circuits can be achieved. GST provides a dramatic change in refractive index upon transition from the amorphous to crystalline state, which is exploited to reversibly control both the extinction ratio and resonance wavelength of the microcavity with an additional gating port in analogy to optical transistors. Our analysis shows excellent sensitivity to the degree of crystallization inside the GST, thus providing the basis for non-von Neuman neuromorphic computing.**

The ability to write, store and retrieve data is at the very heart of information processing. Various techniques are employed to efficiently cope with the vast spread of speed and long term storage needs. In particular Phase Change Memories (PCMs), promise to revolutionize the field of information processing by bridging the gap between the short-term, but very quick operation of on-chip memories and the long-term, but relatively slow storage systems such as solid-state devices and hard-drives [1–3]. Not only can phase change materials switch in a matter of picoseconds [4–6], they are also able to retain information for very long periods of time [2,7]. In addition, they scale extremely well to the nanoscale, with present-day demonstrations of 6nm cells employing electrical switching [7,8]. Specifically $Ge_2Sb_2Te_5$ (GST) is the most commonly used alloy for such applications. By reversibly transforming the crystalline structure between amorphous and crystalline states using electrical pulses, the resistive properties of the thin film can be varied by several orders of magnitude [9]. PCMs also demonstrate a large difference in reflectivity upon phase-transition, an effect that has led to their commercial use in optical storage discs, such as DVDs and Blue-Ray discs [10].

Herein, we propose a chalcogenide-based integrated photonic memory element, with the ability for sub-nanosecond reading and writing, while still retaining data for several years. We analyze the photonic architecture as illustrated in Fig.1(a), which comprises a microring resonator coupled to nanophotonic waveguides. In contrast to photonic memories and mechanical resonators [11], the photonic circuit allows for static tunability which is maintained when the control light has been switched off. We base our analysis on silicon nitride-on-insulator substrates for broadband optical applications. Silicon nitride can be used to fabricate high-quality nanophotonic components for both telecoms and visible applications [12,13]. The feeding waveguide is optimized for single mode operation at 1550nm input wavelength. Inside the ring resonator a small region of the waveguide is suspended similar to waveguides used for nanomechanical sensing [14] and opto-mechanical operation [15]. A thin film of phase change material (GST) is deposited on this suspended region, and forms the basis of the non-volatile memory element of this device. The use of phase change materials, particularly GST, is explored as these materials undergo reversible transformations between crystalline and amorphous states upon application of energy. We consider suspended waveguides to improve thermal isolation of the switching region to ensure that the GST reaches a temperature sufficient to lead to heating-induced crystallization/ amorphization with a short thermal time constant. In order to control the temperature in the suspended section optically a second control port is coupled evanescently to the GST section, where a separate tuning light can be applied independently of the probe light transmitted through the feeding waveguide as shown in Fig.1(b). By varying the coupling gap between the control port and the ring resonator the degree of heat transfer into the GST thin film can be geometrically defined.

---

[*] Email: wolfram.pernice@kit.edu

Light travelling inside the ring resonator couples evanescently to the GST. By changing the crystallization of the GST both the refractive index of the GST and the material absorption vary by a large margin. This significantly affects the modal profile of the propagating mode as shown in the calculated results in Fig.1(c) using COMSOL Multiphysics. The transition from the crystalline to the amorphous state induces significant variation in the modal absorption of the material which directly affects the performance of the optical resonator. When the GST is in the amorphous state the modal profile resembles an optical mode confined largely into the silicon nitride portion of the waveguide. Due to low material absorption in this state, the optical mode is thus largely guided inside the silicon nitride and hence propagates without plasmonic coupling. Upon switching into the crystalline state the material becomes significantly more absorptive and metal-like, leading to stronger plasmonic coupling to the GST thin film (see Fig.1(c), right panel). As a result the mode is confined closer to the metal-like portion, which leads to enhanced optical absorption thus increasing the round-trip loss through the ring resonator.

We analyze the transmission properties of the ring resonator device using coupled mode-theory [16] (CMT) and finite-difference time-domain simulations [17]. Here a micro-ring resonator with a radius of 10μm is considered, which provides a small footprint nanophotonic circuit element. The waveguide cross-section is set to 700x330nm$^2$, corresponding to our previously fabricated waveguide design [18]. The GST covered section is assumed to be 0.5μm long. Assuming typical propagation loss in the silicon nitride waveguide of 3dB/cm as measured previously [18] the chosen parameters provide a compromise between optical Q and free-spectral range. Indeed, in order to obtain wider bandwidth operation it is advantageous to operate the device in a lower Q regime. We initially consider the performance of the device in the amorphous material state. We employ the refractive index profile shown in Fig.2(a) to extract the complex refractive index in dependence of wavelength. For use with the FDTD method the refractive index profile is modeled as a multi-pole Lorentzian material over the wavelength range of interest around 1550nm [19]. In Fig.2(a) the markers denote the data taken from [20], while the solid lines represent the multi-Loretzian fit in good agreement with the measured values. Knowing the roundtrip loss $\alpha_{dB}$ in units of dB/cm, the optical quality factor can then be estimated as $Q = 10\log_{10} e \cdot 2\pi n_g / \lambda \alpha_{dB}$ [21]. From the CMT-model the transmission in the through port shows characteristic resonances with a free-spectral range (FSR) of 48.2nm as presented in Fig.2(b) for a critically coupled device. From the CMT-simulations we estimate an optical quality factor of 9400, equivalent to a linewidth of 160pm. As shown in Fig.2(c), the CMT estimations are also confirmed by the FDTD calculations, as indicated by the overlaid markers in Fig.2(c).

When tuning the probe wavelength transmitted through the feeding waveguide onto resonance, the transmitted signal in the output port is strongly attenuated. Upon switching into the crystalline state, the round-trip loss inside the ring resonator is strongly increased due to the enhanced material absorption in the GST. As a result both the coupling and resonance condition of the resonator are modified significantly. Because the round-trip losses have increased, the existing coupling gap leads to an undercoupled ring resonator, thus increasing the transmission in the through port. This is illustrated in Fig.3(a), where we show the dependence of the device performance on both wavelength and the degree of crystallization. The degree of crystallization $\sigma$ is defined as the volumetric proportion of crystalline material inside the GST, where $\sigma = 0$ corresponds to an entirely amorphous GST and $\sigma = 1$ to a fully crystalline structure. When the round-trip loss is increased, not only the transmission in the through port, but also the width of the resonance increases. For higher round-trip loss stronger coupling into the ring resonator would be required, corresponding to a small coupling gap. Because the gap was optimized for critical coupling for the device with lower roundtrip loss, the larger gap leads thus to a weakly coupled device when the GST has fully switched into the crystalline state. This is illustrated further in the cross-sectional plots in Fig.3(b), which are taken at the intersection dashed line in Fig.3(a). By switching between the amorphous and crystalline state, the transmission past the resonator can be varied between 0 and 90%, thus providing extinction ratio of more than 10dB. By controlling the degree of crystallization we are able to continuously tune the transmission behavior of the device between the low- and high transmission operation. By assigning memory levels to chosen threshold levels of the transmission curve it is thus possible to discriminate between multiple levels of GST crystallization.

We then compute the power of the laser light in the control port required to melt the GST. When the memory cell is operated, the state is readout with a weak probe light placed on the ring resonance, which does not disturb the state of the GST. The control port on the other hand is employed to deliver high intensity optical pulses, which transform the phase composition of the GST. The complex refractive index of the material is related to the material absorption. We assume that the absorbed light is converted to heat inside the waveguide. Because the material absorption of GST is much higher in the visible wavelength range, we employ control light at 700nm to perform the heating operation. From the intensity profile we are able to extract the thermal properties of the system by solving the transient heat transfer equation as a result of conduction, where the thermal material parameters for GST are taken from [22]. The structure is excited with an optical pulse of 600fs width. From the numerical analysis we find that at writing power of less than 5.4pJ, we can raise the temperature inside the released waveguide section to 400 degrees and therefore transform the GST into a crystalline state.

Reversibility of the device requires reamorphization of the crystalline GST, a process that depends on the fall times of the heating pulse – faster fall times are required for reamorphization, which requires efficient heat dissipation. For reamorphization the GST is heated to a higher temperature than for crystallization and rapidly quenched [1]. These phase change processes are intrinsically very fast and the ultimate speed of the device will be determined by the thermal relaxation time.. The thermal relaxation time is extracted by raising the temperature on the beam artificially and then determining the ringdown time in a transient analysis. A snapshot of the temperature distribution of a ringdown simulation is shown in Fig.4(a), where we show the temperature distribution after 3ns from an initial high temperature regime at 300K. As shown in the transient data in Fig.4(b) for a film thickness of 5nm we find a ringdown time of 547ps, within the required range for reamorphization. This is further illustrated in the inset of Fig.4(b), where we plot the temperature evolution in the GST under optical excitation with a 600fs pulse. Because the pulse width is on the order of the thermal ringdown time, the GST is heated very efficiently and subsequently cools down within a sub-nanosecond time interval. From the above analysis we estimate that multi-GHz memory operation of the devices is possible.

In conclusion we have introduced an all-optically tunable photonic platform based on phase-change materials coupled to nanophotonic integrated circuits. We show that multilevel recording can be achieved in such integrated photonic circuits, thereby paving the way for not only ultra-dense photonic memories, but also for all-optical computing using non Von Newman circuits. Furthermore, the ability to switch the PCM section reversibly between separate states also offers the potential for all-optical modulation in an integrated platform.

We thank CD Wright for technical discussions and suggestions. W.H.P. Pernice acknowledges support from the DFG under grant PE 1832/1-1. H. Bhaskaran acknowledges support from EPSRC under grant EP/J00541X/1 and the University of Exeter.

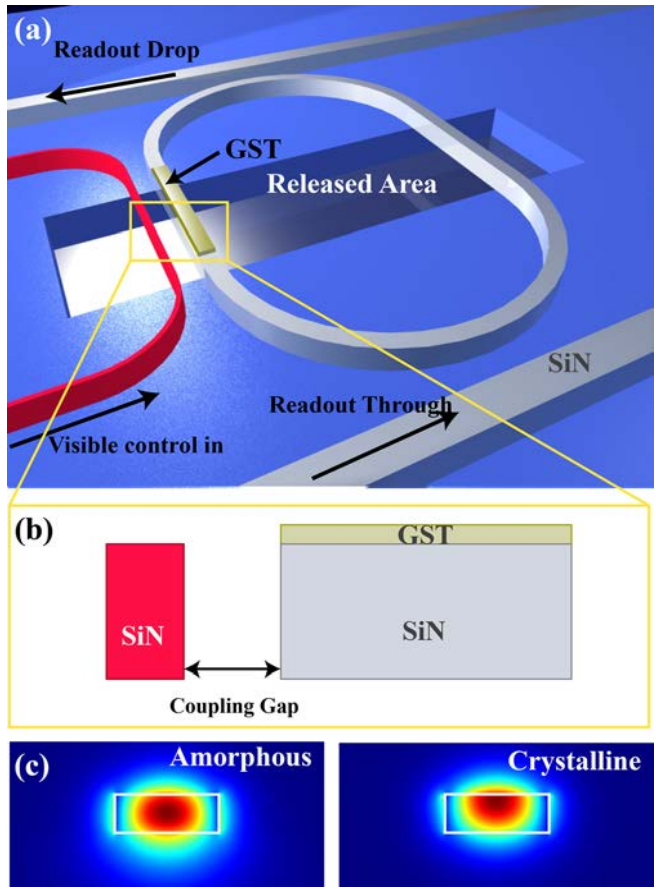

Fig. 1. (a) Schematic overview of the proposed memory element. Light from a control port (red) is coupled evanescently to the ring resonator to perform the switching operation of the GST through photothermal heating. (b) A cross-sectional view of the coupling region showing the control port on the left side and the GST covered free-standing waveguide section on the right side. (c) The calculated modal profiles of the GST covered waveguide cross-section when the GST is in the amorphous state (left panel) and the crystalline state (right panel).

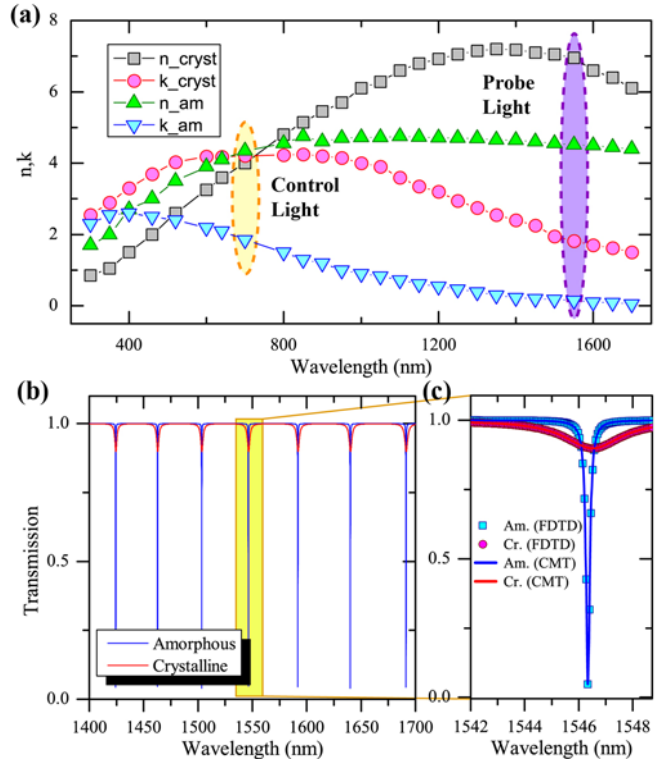

Fig.2. (a) The refractive index of GST in dependence of wavelength used during the CMT and FDTD modeling. Markers are taken from [23], while the solid lines represent the multi-Lorentzian fit. The relevant wavelengths for the control light (orange) and the probe light (purple) are marked. (b) The calculated transmission spectrum of a ring resonator critically coupled in the amorphous state (blue curve). Upon switching the GST into the crystalline state the transmission profile changes significantly (red curve). (c) Zoom into a resonance at 1546nm, showing optical Q of 9400.

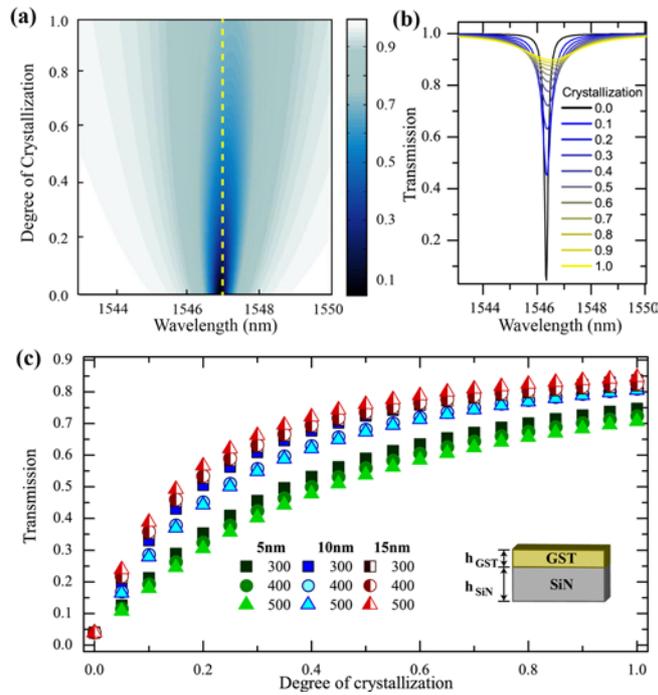

Fig.3. (a) The calculated transmission past the ring resonator in dependence of wavelength and the degree of crystallization of the GST. By fully changing the state of the GST layer, the coupling condition into the ring is switched from the critically coupled to the undercoupled regime. (b) The transmission at the cross-sectional line in (a), showing the increase in transmission as the ring shifts into the weakly coupled regime. (c) The transmission on resonance in dependence of the layer thickness of both GST and Silicon Nitride, as well as the degree of crystallization.

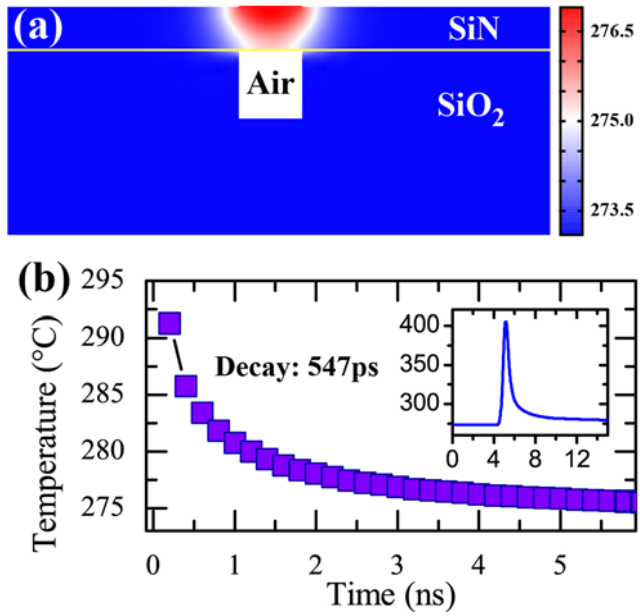

Fig.4 (a) The thermal profile extracted from the calculated intensity distribution of 700nm input light. Shown is the temperature distribution along the center of the free-standing waveguide after 3ns ringdown time. (b) The ringdown time of the GST covered Silicon Nitride beam with a decay constant of 547ps. Inset: Transient simulation of the heating profile within the GST layer using a 600fs optical pulse.